\documentclass[journal,twocolumn]{IEEEtran}
\usepackage{booktabs,algorithm,algpseudocode,subfigure,url, multirow,array}
\usepackage{graphics,epstopdf}

\usepackage[dvips]{color}
\usepackage{amsfonts, amsmath, mathrsfs, amsbsy, amssymb, dsfont, color}
\usepackage{graphicx}

\newcommand\prefixtext[1]{%
  \ifvmode\else\\\@empty\fi
  \noalign{%
    \penalty0%
    \vbox{\mathstrut}%
    \penalty10000%
    \vskip-\baselineskip
    \penalty10000%
    \vbox to 0pt{%
      \normalbaselines
      \ifdim\linewidth=\columnwidth
      \else
        \parshape\@ne
        \@totalleftmargin\linewidth
      \fi
      \vss
      \noindent#1\par}%
      \penalty10000%
      \vskip-\baselineskip}%
      \penalty10000}

\newtheorem{theorem}{Theorem}[section]

\newtheorem{proposition}{Proposition}[section]

\newcommand{\qed}{\nobreak \ifvmode \relax \else
      \ifdim\lastskip<1.5em \hskip-\lastskip
      \hskip1.5em plus0em minus0.5em \fi \nobreak
      \vrule height0.75em width0.5em depth0.25em\fi}

\DeclareMathAlphabet{\mathpzc}{OT1}{pzc}{m}{it}
\newcommand{\comment}[1]{}
\def\Re{\mathrm{Re}}
\def\Im{\mathrm{Im}}
\def\CRB{\mathrm{CRB}}
\def\zero{\boldsymbol{0}}
\def\bal{\pmb{\mathscr{\alpha}}}
\def\bbe{\boldsymbol{\beta}}

\def\bth{\boldsymbol{\theta}}

\def\bom{\boldsymbol{\omega}}

\def\tr{\text{tr}}
\def\({\left(}
\def\){\right)}
\def\[{\left[}
\def\]{\right]}
\def\BEq{\begin{eqnarray}}
\def\EEq{\end{eqnarray}}
\def\BE*{\begin{eqnarray*}}
\def\EE*{\end{eqnarray*}}
\def\BA{\begin{array}}
\def\EA{\end{array}}

\def\0{\mathbf{0}}
\def\1{\mathbf{1}}
\def\a{\mathbf{a}}
\def\A{\mathbf{A}}

\def\bfb{\mathbf{b}}
\def\B{\mathbf{B}}

\def\bC{\mathbb{C}}

\def\D{\mathbf{D}}

\def\bE{\mathbb{E}}

\def\F{\mathbf{F}}

\def\g{\mathbf{g}}
\def\G{\mathbf{G}}

\def\I{\mathbf{I}}

\def\M{\mathbf{M}}

\def\p{\mathbf{p}}

\def\R{\mathbf{R}}

\def\bR{\mathbb{R}}
\def\s{\mathbf{s}}

\def\bfu{\mathbf{u}}

\def\bfv{\mathbf{v}}
\def\V{\mathbf{V}}
\def\w{\mathbf{w}}

\def\x{\mathbf{x}}
\def\X{\mathbf{X}}
\def\y{\mathbf{y}}

\def\z{\mathbf{z}}

\def\and{\prefixtext{and}}

\def\diag{{\rm diag}}

\allowdisplaybreaks

\newtheorem{remark}{Remark}[section]
\newtheorem{claim}{Claim}[section]

\graphicspath{{Figures/}}

\begin{document}

\title{Phase Retrieval Using Feasible Point Pursuit: Algorithms and Cram\'er-Rao Bound}
\author{Cheng Qian, \emph{Student Member, IEEE,} Nicholas D. Sidiropoulos, \emph{Fellow, IEEE}, Kejun Huang, \emph{Student Member, IEEE,} Lei Huang, \emph{Senior Member, IEEE} and H. C. So, \emph{Fellow, IEEE}
\thanks{Preliminary conference version of part of this work appears in \cite{QSHHS:ICASSP2016}.
C. Qian is with the Department of Electronics and Information Engineering, Harbin Institute of Technology, Harbin, 150001 China (e-mail: qianc@umn.edu).
N. D. Sidiropoulos and K. Huang are with the Department of Electrical and Computer Engineering, University of Minnesota, Minneapolis, MN 55455 USA (e-mail: nikos@umn.edu, huang663@umn.edu).
L. Huang is with the College of Information Engineering, Shenzhen University, Shenzhen, 518060 China (e-mail: dr.lei.huang@ieee.org).
H. C. So is with the Department of Electronic Engineering, City University of Hong Kong, Hong Kong, China (e-mail: hcso@ee.cityu.edu.hk).
The work of N. Sidiropoulos was supported by NSF CIF-1525194. K. Huang was supported by a UMII dissertation fellowship. C. Qian is on leave from the Department of Electronics and Information Engineering, Harbin Institute of Technology, China, supported in part by the Natural Science Foundation of China (NSFC) under Grant No. 61171187 and the Chinese Scholarship Council.
}
}
\maketitle

\begin{abstract}
Reconstructing a signal from squared linear (rank-1 quadratic) measurements is a challenging problem with important applications in optics and imaging, where it is known as \emph{phase retrieval}. This paper proposes two new phase retrieval algorithms based on non-convex quadratically constrained quadratic programming (QCQP) formulations, and a recently proposed approximation technique dubbed \emph{feasible point pursuit (FPP)}. The first is designed for uniformly distributed bounded measurement errors, such as those arising from high-rate quantization (B-FPP). The second is designed for Gaussian measurement errors, using a least squares criterion (LS-FPP). Their performance is measured against state-of-the-art algorithms and the Cram\'er-Rao bound (CRB), which is also derived here. Simulations show that LS-FPP outperforms the existing schemes and operates close to the CRB. Compact CRB expressions, properties, and insights are obtained by explicitly computing the CRB in various special cases -- including when the signal of interest admits a sparse parametrization, using harmonic retrieval as an example.
\end{abstract}
\begin{IEEEkeywords}
	Phase retrieval, quadratically constrained quadratic programming (QCQP), semidefinite programming (SDP), feasible point pursuit (FPP), Cram\'er-Rao bound (CRB).
\end{IEEEkeywords}

\section{Introduction}
Phase retrieval is the problem of reconstructing a signal $\x\in\bC^{N}$ from measurements of the form
\begin{align}\label{model}
  y_i = |\a_i^H\x|^2, \quad i\in\left\{1,\cdots,M\right\}
\end{align}
where $|\cdot|$ is the magnitude of a complex number, $(\cdot)^H$ is the conjugate transpose and $\a_i\in\bC^N$ is a known measurement vector.
The above problem appears in many applications such as crystallography \cite{S1:1}, diffraction imaging \cite{S1:2}-\cite{S1:3} and microscopy \cite{S1:4}-\cite{S1:5}, where it is often far easier to measure the magnitude than the phase.

During the past decades, numerous phase retrieval solvers have been developed in the literature. Among them, the Gerchberg-Saxton (GS) \cite{AP:1} and Fienup \cite{AP:2} algorithms are the most well-known and widely used methods in practice. These approaches are based on alternating optimization in which the unknown $\x$ is iteratively estimated by solving a least squares (LS) problem, i.e.,
\begin{equation}\begin{aligned}
  \min_{\x,\bfu ~|~ |u_i|=1, ~\forall i}\ ||\sqrt{\y}\odot\bfu-\A^H\x||_2^2
\end{aligned}\end{equation}
where $\y = [y_1\ \cdots\ y_M]^T$ is the data vector, $\A=[\a_1\ \cdots\ \a_M]$ is the known measurement matrix, $\bfu$ is the phase of $\sqrt\y$ (an extra unknown, together with $\x$), $||\cdot||_2$ is the 2-norm and $\odot $ denotes element-wise multiplication. The main problem with this type of algorithms is that they tend to hit local minima, thus requiring careful initialization, and often fail to perform satisfactorily even after multiple initializations.

Recently, modern convex relaxation techniques were applied to phase retrieval. \textit{PhaseLift} \cite{C1}-\cite{C2} employs matrix lifting to recast phase retrieval as a semi-definite programming (SDP) problem. Specifically, the PhaseLift scheme regards the measurements in \eqref{model} as a linear function of $\X=\x\x^H$ which is a rank-1 Hermitian matrix, i.e.,
\begin{align}
  y_i = |\a^H_i\x|^2 = \x^H\a_i\a_i^H\x = \tr(\A_i\X)
\end{align}
where $\A_i = \a_i\a_i^H$ and $\tr(\cdot)$ denotes the trace of a matrix.
Thus, the recovery of $\x$ is equivalent to finding a positive semidefinite rank-1 matrix $\X$ through solving a rank minimization problem:
\begin{equation}\label{s1:pl}\begin{aligned}
  \min_{\X}&\ \text{rank}(\X) \\
  \text{s.t. }&\ y_i = \tr(\A_i\X),\ i\in\left\{1,\cdots,M\right\} \\
        &\ \X\succeq0.
\end{aligned}\end{equation}
Since rank minimization is a non-convex problem which is difficult to solve in a computationally efficient manner, PhaseLift relaxes \eqref{s1:pl} via semidefinite relaxation (SDR) -- see \cite{luo} for a tutorial overview. It has been shown in \cite{C1} that if the measurement vectors are i.i.d. Gaussian distributed, then PhaseLift can recover $\x$ with high probability when the number of measurements $M\sim\mathcal{O}(N\log N)$. However, when the measurements are corrupted by noise, there is no guarantee that PhaseLift will yield a rank-1 solution \cite{Eldar}.

\textit{PhaseCut} \cite{PC} takes a similar approach as PhaseLift, but instead of directly aiming for $\x$ it tries to find $\bfu$ first. Substituting the conditional LS estimate $\hat\x = (\A^H)^\dagger\diag(\sqrt\y)\bfu$ of $\x$ given $\bfu$ where $\diag(\cdot)$ denotes a diagonal matrix and $(\cdot)^\dagger$ denotes the pseudo-inverse, PhaseCut aims at recovering $\bfu$ by solving the non-convex quadratic program
 \begin{equation}\label{s1:pc}\begin{aligned}
  \min_{\bfu}&\ \bfu^H\M\bfu \\
  \text{s.t. }&\ |u_i| = 1, i\in\left\{1,\cdots,M\right\}
 \end{aligned}\end{equation}
where $\M = \diag(\sqrt\y)(\I_M-\A^H(\A^H)^\dagger)\diag(\sqrt\y)$ with $\I_M$ being a $M\times M$ identity matrix. Formulation \eqref{s1:pc} resembles the classical \emph{MaxCut} problem in networks, enabling fast semidefinite relaxation algorithms originally developed for MaxCut to be adapted for PhaseCut. This makes PhaseCut faster than PhaseLift.

More recently, a new approach to phase retrieval was proposed, in what appears to be an instance of a new algorithmic genre that relies on smart `statistical' initialization followed by relatively simple descent-type refinement named \textit{Wirtinger Flow (WF)} \cite{C3}. It has been theoretically shown that when sufficiently many i.i.d. Gaussian measurement vectors are used, WF will recover the desired solution with high probability. However, recovery cannot be guaranteed when the number of measurements is small, or when the measurement vectors are not random -- mainly because the principal eigenvector used for initialization is not a good approximation of $\x$ in such cases. This means that for systematic (non-random) measurement designs and/or relatively short sample sizes there is considerable room for improvement.

In this paper, the focus is on recovering $\x$ from noisy measurements, i.e.,
\begin{align}\label{model1}
  y_i = |\a_i^H\x|^2 + n_i, \quad i\in\left\{1,\cdots,M\right\}
\end{align}
where $n_i$ is additive noise. To this end, in Section II, two novel algorithms are developed. These algorithms build upon a method called \emph{feasible point pursuit (FPP)} that we recently developed for non-convex quadratically constrained quadratic programming (QCQP) problems \cite{FPP}. The first algorithm (B-FPP) is designed for independent and uniformly distributed bounded measurement errors, such as those arising from high-rate quantization. The second (LS-FPP) is designed for i.i.d. Gaussian measurement errors, thereby using a LS criterion. Their performance is measured against state-of-art algorithms and the general Cram\'er-Rao bound (CRB) for phase retrieval from magnitude measurements in additive Gaussian noise, which is also derived here in terms of phase and amplitude of the input signal. Interestingly, only partial CRB results under additional model restrictions and/or different noisy measurement models (e.g., for real- and complex-valued $\x$ \cite{crb_pr2}-\cite{crb_pr4}, noise added prior to taking the magnitude \cite{crb_pr1}, 2-D Fourier-based measurements \cite{crb_pr0}) were previously available, despite decades of research in phase retrieval. Simulations show that LS-FPP outperforms the state-of-art and operates close to the CRB. Compact CRB expressions, properties, and insights are obtained by simplifying the CRB in special cases. These can help improve the design of measurement apparatus, by providing a way to score different designs.

Section IV presents a special case where $\x$ is in the form of a linear combination of several Vandermonde vectors, i.e., a harmonic mixture, leading to harmonic retrieval from rank-1 quadratic measurements. By predefining an overcomplete frequency basis, sparsity in the frequency domain can be exploited, resulting in modified versions of B-FPP and LS-FPP for sparse phase retrieval. Furthermore, the CRB for frequency estimation is derived for this case.

Section V contains numerical simulations designed to illustrate the performance of the proposed algorithms versus PhaseCut, PhaseLift, WF, and CRB. Finally, conclusions are drawn in Section VI.

\section{Proposed Algorithms}
In this section, we formulate the phase retrieval problem as non-convex QCQP in two different ways, and derive two corresponding algorithms, B-FPP and LS-FPP, to recover $\x$.
\subsection{B-FPP Algorithm}
In the absence of noise, phase retrieval can be cast as
\begin{equation}\label{FPP1}\begin{aligned}
  \min_\x\quad &||\x||^2_2 \\
  \text{s.t.}\quad &\x^H\A_i\x = y_i,\ i \in \left\{1,\cdots, M\right\},
\end{aligned}\end{equation}
i.e., a minimum norm solution to a system of quadratic equations in $\x$. If the equality constraints are consistent, then using the minimum norm to pick a solution can be motivated from a Bayesian perspective, if we assume a zero-mean uncorrelated complex circularly symmetric Gaussian prior on $\x$.\footnote{Since $||\x||^2_2=\tr(\x\x^H)$, the minimum norm criterion is also reminiscent of semidefinite relaxation of rank minimization.} In practice noise will render the equality constraints in \eqref{FPP1} inconsistent, so \eqref{FPP1} will not admit any solution. High-resolution uniform scalar quantization of otherwise noiseless quadratic measurements will result in additive quantization noise that is independent across measurements, bounded, and approximately uniformly distributed over the quantization interval. This motivates using interval constraints, as follows:
\begin{align}\label{s2:FPP_Cons}
  |y_i - \x^H\A_i\x|\leq \epsilon,\ \forall i.
\end{align}
Replacing the constraints in \eqref{FPP1} by \eqref{s2:FPP_Cons} yields
\begin{subequations}\label{FPP2}\begin{align}
  \min_\x\quad &||\x||^2_2 \\
  \text{s.t.}\quad &\x^H\A_i^H\x\leq y_i+\epsilon \\
  &\x^H\A_i^H\x\geq y_i-\epsilon,\forall i. \label{FPP:13}
\end{align}\end{subequations}
It is clear that due to the non-convex constraints in \eqref{FPP:13}, \eqref{FPP2} belongs to the class of non-convex QCQP problems which is NP-hard in general. For $\epsilon=0$ we recover the `standard' phase retrieval problem, which is NP-hard \cite{PRNP}.

To approximately solve \eqref{FPP2}, we follow \cite{FPP}. Recall that $\A_i$ is of rank one and it has only one positive eigenvalue. For any $\z$ and $\x$, we have
\begin{align}\label{FPP3}
  (\x-\z)^H\A_i(\x-\z)\geq 0.
\end{align}
Expanding the left-hand side of \eqref{FPP3} yields
\begin{align}
  \x^H\A_i\x \geq 2\Re\{\z^H\A_i\x\} - \z^H\A_i\z
\end{align}
where $\Re\left\{\cdot\right\}$ takes the real part of its argument.
Following the rationale in \cite{FPP}, we replace \eqref{FPP:13} by
\begin{align}\label{FPP4}
  2\Re\{\z^H\A_i\x\} + s_i \geq \z^H\A_i\z + y_i-\epsilon
\end{align}
where $s_i\geq 0$ is a slack variable. The idea here is that linear restriction turns the non-convex problem into a convex one, but at the risk of infeasibility. The slack variables restore feasibility, but they should be sparingly used \cite{FPP}. This leads to the following formulation:
\begin{equation}
          \begin{aligned}
            \min_{\x,\s}\quad & ||\x||^2_2 + \lambda\sum_{i=1}^M s_i \\
            \text{s. t.}\quad & \x^H\A_i\x\leq y_i + \epsilon\\
            &2\Re\{\z^H\A_i\x\} + s_i \geq \z^H\A_i\z + y_i-\epsilon \\
            &s_i\geq 0,\ \forall i
          \end{aligned}\label{Q1}
\end{equation}
where $\s = [s_1\ \cdots\ s_M]^T$ and the regularization parameter $\lambda$ balances the original cost versus the slack penalty term. Starting with an initial (possibly random) $\z$, we solve a sequence of problems of type \eqref{Q1} to obtain $(\x_k,\s_k)$, and setting $\z_{k+1}=\x_k$. Since the cost function in \eqref{Q1} is independent of $k$ and the solution of the $k$th iteration is also feasible for the $(k+1)$th iteration, this will always return a non-increasing cost sequence \cite{FPP}. In other words, the optimal value of the cost function in each iteration step is non-increasing. It follows that this sequential process will converge in terms of the cost function. The steps for B-FPP are summarized in \textbf{Algorithm \ref{FPP}}.
\begin{algorithm}
   \caption{B-FPP Algorithm for Phase Retrieval}\label{FPP}
    \begin{algorithmic}[1]
      \Function{$\hat\x=$ B-FPP}{$\A,\y,\lambda,\epsilon,\z$}

	\Repeat

        \State $\hat\x\gets$ solution of \eqref{Q1}
		\State $\z = \hat\x$
		
	\Until{a stopping criterion on the cost function of \eqref{Q1} is satisfied}

     \EndFunction

\end{algorithmic}
\end{algorithm}

\begin{figure}[h!]
\centering
\includegraphics[width=7.2cm]{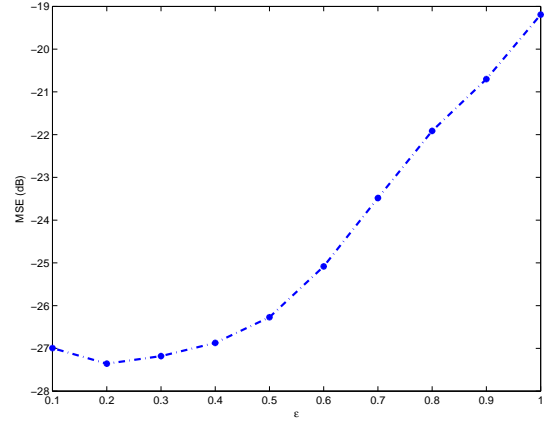}
\caption{MSE versus $\epsilon$. ($N=16$, $M=80$)}
\label{fig:FPP_Epsilon}
\end{figure}

Whereas B-FPP has been motivated from a uniform high-resolution quantization point of view (and indeed matches that noise model), the resultant algorithm can also be used for Gaussian noise, although the choice of $\epsilon$ is less obvious in this case. It is instructive to illustrate this by means of an example. Assume $\x$ is uncorrelated zero-mean Gaussian with length $N=16$, and $M=80$ measurements are used for signal recovery. 200 Monte-Carlo trials are employed to calculate the mean square error (MSE). In each trial, $\a_i$, $\x$ and $\sigma_n$ are fixed, and the noise is generated from a white Gaussian process with mean zero and standard deviation $\sigma_n=0.4$. Fig. \ref{fig:FPP_Epsilon} shows the MSE versus $\epsilon$. It is observed that when $\epsilon<0.4$, B-FPP exhibits a relative small MSE. Otherwise, its performance gets worse as $\epsilon$ increases. We conclude that B-FPP still works for Gaussian noise, provided $\epsilon \sim \sigma_n$.

\subsection{LS-FPP Algorithm}
The B-FPP method requires a user-defined tolerance $\epsilon$ to bound the noise perturbation in the constraints, which is difficult to appropriately determine from the magnitude measurements without prior knowledge of the noise standard deviation. More to the point, B-FPP is not tailored for Gaussian noise. In this section we develop \textit{LS-FPP} based on the LS criterion, which is equivalent to maximum likelihood for additive white Gaussian noise. The LS formulation of phase retrieval has been recently considered in \cite{C3}, but the WF approach does not always work well, as we will show in our simulations in Section V. This is not surprising, of course, since we are dealing with an NP-hard problem. Our contribution here is to recast LS phase retrieval as a non-convex quadratic-plus-linear problem, and then approximate it using FPP. As we will show, our approach gives consistently better approximation results, especially in challenging scenarios, at the cost of additional computational complexity.

The LS formulation for phase retrieval is \cite{C3}
\begin{align}\label{LS}
  \min_\x\  \sum_{i=1}^M(y_i-\x^H\A_i\x)^2
\end{align}
The first step in our approach is to recast \eqref{LS} in the following equivalent form
\begin{equation}\label{NonFPP:1}
  \begin{aligned}
    \min_{\w,\x}\quad &||\w||_2^2\\
    \text{s. t.}\quad & \x^H\A_i\x + w_i = y_i,\ \forall i
  \end{aligned}
\end{equation}
where
\begin{align}
	\w=[w_1\ \cdots\ w_M]^T
\end{align}
with $(\cdot)^T$ being the transpose.
We rewrite the equality constraints as
\begin{subequations}\begin{align}
	\x^H\A_i\x + w_i \leq&\ y_i \label{c1}\\
	\x^H\A_i\x + w_i \geq&\ y_i. \label{cons2}
\end{align}\end{subequations}

In a similar manner as we process the non-convex constraints in FPP, \eqref{cons2} can be replaced by
\begin{align}\label{c2}
  2\Re\{\z^H\A_i\x\}  + w_i + s_i\geq y_i + \z^H\A_i\z
\end{align}
to obtain the following convex QCQP:
\begin{equation}
  \begin{aligned}
    \min_{\x,\w,\s}\quad &||\w||^2_2 + \lambda\sum_{i=1}^M s_i\\
    \text{s. t.}\quad & 2\Re\{\z^H\A_i\x\}  + w_i + s_i\geq y_i + \z^H\A_i\z\\
    & \x^H\A_i\x + w_i \leq y_i, \\
    & s_i\geq 0, \forall i.
  \end{aligned}\label{Q2}
\end{equation}

The steps for LS-FPP are summarized in \textbf{Algorithm \ref{LS-FPP}}.
\begin{algorithm}
   \caption{LS-FPP Algorithm for Phase Retrieval}\label{LS-FPP}
    \begin{algorithmic}[1]
      \Function{$\hat\x=$ LS-FPP}{$\A,\y,\lambda,\z$}

        \Repeat

         \State $\hat\x\gets$ solution of \eqref{Q2}
         \State $\z = \hat\x$

        \Until{a stopping criterion on the cost function of \eqref{Q2} is satisfied}

       \EndFunction
\end{algorithmic}
\end{algorithm}

Some important remarks are in order:

\noindent $\bullet$ The problems in \eqref{Q1} and \eqref{Q2} are convex and can be solved via interior point methods \cite{InteriorPointMethod1}-\cite{InteriorPointMethod2}. The worst-case complexity of solving \eqref{Q1} and \eqref{Q2} are $\mathcal{O}\big((N+2M)^{3.5}\big)$ and $\mathcal{O}\big((N+3M)^{3.5}\big)$, respectively. Moreover, few outer iterations of B-FPP or LS-FPP are usually needed, so that the overall approximation is often manageable for moderate $N$.

\noindent $\bullet$ In both B-FPP and LS-FPP, the regularizer $\lambda$ is chosen according to \cite{FPP}, where it is suggested to use $\lambda\gg1$ to steer the iterates towards the feasible region. Our experience is that FPP is not very sensitive to the choice of $\lambda$. Usually, $\lambda=10$ works well for B-FPP and LS-FPP in most scenarios.

\noindent $\bullet$ Invoking \cite[Theorem~1]{meisam_thesis}, it follows that Algorithms ~\ref{FPP} and \ref{LS-FPP} have a convergent subsequence. If it happens that the slack variable $\s$ at the limit point is zero, then from \cite[Theorem~1]{meisam_thesis} it follows that the $\x$ variable at the limit point is also a KKT point of the original problem (\ref{FPP2}) or (\ref{NonFPP:1}), respectively. Given the NP-hard nature of (\ref{FPP2}) and (\ref{NonFPP:1}), these convergence claims may be reassuring; but it is important to not lose sight of the following caveat. Whereas numerical experiments suggest that if the original problem is feasible then $\s$ is very likely to be zero at the limit point, this is not always true -- counterexamples have been found \cite{FPP},  and this is consistent with the fact that the feasibility problem is NP-hard.

\noindent $\bullet$ Our work was inspired by the FPP-SCA (successive convex approximation) algorithm originally proposed for general non-convex QCQPs in \cite{FPP}. The idea behind the algorithm is closely related to the well-known \emph{difference of convex programming} (DCP) and the \emph{convex-concave procedure} (CCP) in optimization. The difference is these classical procedures assume the availability of a feasible starting point, which is the core challenge in our context. FPP can be interpreted as first adding slack variables and a slack penalty to the original problem to ensure feasibility (thereby circumventing the initialization challenge), followed by application of DCP/CCP to the augmented problem, see \cite{FPP}. The same idea was independently proposed in a parallel submission which appeared later in \cite{ccp1}. An early version of the same basic idea can be found in \cite{ccp2}, which however neither considered general QCQPs, nor did it demonstrate that the method works well, especially relative to standard semidefinite relaxation and randomization baselines.

Given the apparent success of FPP in solving challenging QCQP problems, we therefore propose using FPP to solve the phase retrieval problem, where feasibility is the key stumbling block. Whereas optimization theory measures success via the optimality gap in terms of the cost function, estimation theory naturally focuses on the estimation error. We therefore need a statistical estimation baseline to assess how well FPP works when applied to phase retrieval.

\section{Cram\'er-Rao Bound for Phase Retrieval}
In this section, we derive the CRB for phase retrieval for measurements contaminated by additive white Gaussian noise {\em after} magnitude squaring.

\subsection{Previous Work on CRB}
Let us summarize the (surprisingly scant) prior work on the CRB for phase retrieval. Balan \cite{crb_pr3} has derived the Fisher Information Matrix (FIM) for the model in \eqref{model1} for complex-valued $\x$. Realizing that the FIM is singular, and implicitly attributing this to the lack of global phase identifiability, he suggested using side information about $\x$ (e.g., assuming one particular component of $\x$ is real-valued) to reduce the dimension of the FIM, resulting in a full-rank matrix. Thus, the CRB can be computed by taking the inverse of the dimension-reduced FIM. Similar results have also been considered in \cite{crb_pr4}, where the last row and column of the FIM are deleted. However, these assumptions are impractical and identifiability neither implies nor is implied by a nonsingular FIM \cite{BasuBre2000}. Instead of making additional assumptions on $\x$ to force a non-singular FIM, we can instead use the pseudo-inverse of the full FIM as a lower bound:
\begin{claim}\label{theorem_crb_c}
	For $\x\in\bC^N$, the CRB matrix for the phase retrieval model in (6) is
	\begin{align}
        \CRB_c = \F_c^\dagger
	\end{align}
	where the FIM is given by
	\begin{align}\label{FIMc}
		\F_c = \frac{4}{\sigma_n^2}\G_c\G_c^T
	\end{align}
	with
	\begin{align}\label{Gc}
		\G_c =
		\begin{bmatrix}
			\Re\{\A\diag\{\A^H\x\}\} \\
			\Im\{\A\diag\{\A^H\x\}\}
		\end{bmatrix}.
	\end{align}
\end{claim}
\begin{IEEEproof}
	The FIM has been derived in \cite{crb_pr3} and \cite{crb_pr4} (in different but equivalent form). When the FIM is rank deficient, its pseudo-inverse is a valid lower bound on the MSE of any unbiased estimator \cite{crb1,crb2}, albeit this bound is generally looser than the usual CRB \cite{crb0}. Perhaps surprisingly, this `optimistic' bound is often attainable in practice and therefore predictive of optimal estimator performance -- see \cite{crb_huang} and our simulations that follow. Strictly speaking, the pseudo-inverse of a singular FIM is not the usual CRB, and some researchers distinguish the two bounds; but this is a technical detail with little practical consequence, so we will refer to the resultant bound as the CRB.
\end{IEEEproof}

In the case of real ${\bf x}$, Balan's result in \cite{crb_pr2} is valid only for real measurement vectors. The CRB for real $\x$ can be easily derived from Theorem \ref{theorem_crb_c}. The result is as follows.
 \begin{claim}\label{theorem_crb_r}
 	For $\x\in\bR^N$, the CRB matrix for the phase retrieval model in (6) is
	\begin{align}
        \CRB_r = \F_r^{-1}
	\end{align}
 	where $(\cdot)^{-1}$ denotes the inverse and
 	\begin{align}\label{FIM_r}
 		\F_r = \frac{4}{\sigma_n^2}\G_r\G_r^T
 	\end{align}
 	with
 	\begin{align}
 		\G_r = \Re\{\A\diag\{\A^H\x\}\}.
 	\end{align}
 \end{claim}
Balan also derived \cite{crb_pr1} the FIM for complex white Gaussian noise added {\em prior} to taking the magnitude square, i.e., $y_i=|\a_i^H\x + n_i|^2$, which is different from our model in \eqref{model}. We also note \cite{crb_pr0}, where the CRB has been derived for a 2-D phase retrieval model with 2-D Fourier measurements.

\subsection{CRB on Phase and Amplitude of $\x$}
The phase of a complex signal is often more informative than its amplitude -- see \cite{Eldar} for a striking illustration. This is particularly true when one is interested in measuring frequency- or phase-modulated signals, where the amplitude carries little (if any) information. This motivates using an explicit amplitude-phase parametrization of the unknown vector, and computing the associated CRB. This is the subject of the next theorem. We also note that many other (non-Gaussian) noise probability density functions possessing everywhere continuous first and second derivatives can be easily handled -- as the corresponding CRB only differs by a noise distribution-specific shape factor \cite{SwamiCRB}.
 \begin{theorem}\label{crb_ap}
 	The CRB for the phase retrieval model in \eqref{model1} on the phase and amplitude of $\x$ is
 	\begin{align}
 		\text{CRB} = \F^\dagger
 	\end{align}
 	where the FIM is given by
 		  \begin{align}\label{FIM_c}
 		      \F = \frac{4}{\sigma_n^2}  \G \G^T
 		  \end{align}
 		  with
 		  \begin{align}
 		  	  \G =
 		  	\begin{bmatrix}
 		  		\Re\left\{\diag(e^{-j\bth})\A\diag(\A^H\x)\right\}\\
 		  		\Im\left\{\diag(\x^*)\A\diag(\A^H\x)\right\}
 		  	\end{bmatrix}.
 		  \end{align}
 	In particular, the CRB for phase and amplitude have closed-form expressions as
 	\begin{align}
 		\text{CRB}_{\theta} =&\ (\F_{\theta\theta} - \F_{\theta b}\F_{bb}^{-1}\F_{b\theta})^\dagger \\
 		\text{CRB}_b=&\ (\F_{bb} - \F_{b\theta} \F_{\theta\theta}^{-1}\F_{\theta b})^\dagger
 	\end{align}
 	where $\F_{\theta\theta} $, $\F_{bb}$, $\F_{\theta b}$ and $\F_{b\theta }$ are defined in \eqref{F111}-\eqref{F444}, respectively. Moreover, the variance on phase and amplitude of any unbiased phase retrieval estimators designed for model \eqref{model1} is bounded below by
 \begin{align}
   \bE\left[\big|\big|\hat\bth-\bth\big|\big|^2_2\right] &\geq \mathrm{trace}(\mathrm{CRB}_\theta) \\
   \bE\left[\big|\big||\hat\x|-|\x|\big|\big|^2_2\right] &\geq \mathrm{trace}(\mathrm{CRB}_b)
 \end{align}
 \end{theorem}
\begin{IEEEproof}
	See Appendix \ref{App_proof_cmplx}.
\end{IEEEproof}

\subsection{Some Useful Properties}
The following proposition shows that the FIM in \eqref{FIM_c} is always singular for nonzero $\x$.
\begin{proposition}
	When $\A$ is nontrivial and has full row rank $N$, for both real and complex $\x$, the FIM $\F$ in \eqref{FIM_c} is always singular with rank deficit equal to one, and for any nonzero $\alpha$, $[\zero_N^T, \alpha\boldsymbol{1}_N^T]^T$ always lies in the null space of $\F$.
\end{proposition}
\begin{IEEEproof}
	See Appendix \ref{proof_of_rank1}.
\end{IEEEproof}

As we have pointed out in Section III-A, for complex $\x$ the FIM in \eqref{FIMc} is always singular. For real $\x$, the FIM in \eqref{FIM_r} is nonsingular. Related observations have been noted in \cite{crb_pr2}-\cite{crb_pr4} but without any proof. We provide precise claims and proofs in the following.
\begin{proposition}
	When $\A$ is nontrivial and has full row rank $N$, for complex-valued $\x$, $\F_c$ is always singular with rank deficit equal to one, and the direction $\big[-\Im\{\x\}^T\ \Re\{\x\}^T\big]^T$ is always in its null space.
\end{proposition}
\begin{IEEEproof}
  See Appendix \ref{proof_of_rank1_2}.
\end{IEEEproof}

\begin{proposition}
	When $\A$ is nontrivial and has full row rank $N$, for real-valued $\x$, $\F_r$ is always nonsingular.
\end{proposition}
\begin{IEEEproof}
	See Appendix \ref{proof_of_fullrank_Fr}.
\end{IEEEproof}

We intuitively expect a reduced bound when more measurements are added. The following theorem shows that this is indeed true.
\begin{proposition}
  For given $\x$ and fixed $\sigma_n$, the CRB in Theorem \ref{crb_ap} decreases as more measurements are made available:
  \begin{align}\label{s2:crbm1}
     \CRB(\A(:,1:M+1)) \preceq \CRB(\A(:,1:M)),
  \end{align}
  where $\A(:,\ell:r)$ is (Matlab notation for) the submatrix of $\A$ comprising columns $\ell$ to $r$ inclusive.
\end{proposition}
\begin{IEEEproof}
  To prove \eqref{s2:crbm1}, we first rewrite $ \G$ as
  \begin{align}
     \G =&
     \begin{bmatrix}
	     \Re\left\{\diag(e^{-j\bth})[\A_1\x\ \cdots\ \A_M\x]\right\}\\
	     \Im\left\{\diag(\x^*)[\A_1\x\ \cdots\ \A_M\x]\right\}
     \end{bmatrix} \notag\\
     =:& \ [\g_1 \ \cdots\ \g_M]
  \end{align}
  where
  \begin{align}
  	\g_i =
  	\begin{bmatrix}
	     \Re\left\{\diag(e^{-j\bth})\A_i\x\right\}\\
	     \Im\left\{\diag(\x^*)\A_i\x\right\}
     \end{bmatrix}.
  \end{align}
  Define $\F(M)$ and $\F(M+1)$ as the FIMs for $M$ and $(M+1)$ measurements, respectively. Then, we have
  \begin{align}\label{s2:FcM1}
    \F(M+1) = \F(M) + \frac{4}{\sigma_n^2} \g_{M+1}\g_{M+1}^T.
  \end{align}
  It is seen that the second term in \eqref{s2:FcM1} is positive semidefinite. By taking the pseudo-inverse of \eqref{s2:FcM1}, \eqref{s2:crbm1} is established straightforwardly.
\end{IEEEproof}

\section{Harmonic Retrieval from Rank-one Quadratic Measurements}
\subsection{Signal Model}
In this section, we consider a special case of (6) when $\x$ is a linear combination of $L$ Vandermonde vectors where each vector contains a single frequency, i.e.,
\begin{align}\label{s4:x}
  \x = \sum_{\ell=1}^L\gamma_\ell\bfv(\omega_\ell)
\end{align}
Here, $\omega_\ell$ and $\gamma_\ell$ stand for the $\ell$th unknown frequency and complex amplitude, respectively, and
\begin{align}
  \bfv(\omega_\ell) = 
  \begin{bmatrix}
    e^{j\omega_\ell} & \cdots & e^{jN\omega_\ell}
  \end{bmatrix}^T.
\end{align}
The main problem here is to estimate the frequencies $\{\omega_1\ \cdots\ \omega_L\}$ from $\y$. Classical line spectra estimators such as MUSIC and ESPRIT assume that $\x$ is sampled directly and there is no phase noise. What if we observe {\em generalized samples}, i.e., linear combinations of the elements of $\x$, and these are subject to phase noise, i.e., $\p = \text{diag}({\bf u}) {\bf A}^H \x$, where $u_i$ models phase noise in the $i$th measurement ($|u_i|=1$), which could arise, e.g., due to phase offsets when different measurements are collected by different sensors in a network sensing scenario. In this case, the phase of $\p$ is clearly uninformative, and we might as well get rid of it by working with $|\p|$ - see also \cite{v2}. This yields a phase retrieval problem where the unknown $\x$ possesses harmonic structure. Can we adapt our algorithms and bounds to account for this structure?

\subsection{Sparse B-FPP and LS-FPP}
We propose to adapt B-FPP and LS-FPP using sparse regression with an overcomplete Vandermonde dictionary. Let $\tilde\V\in\bC^{N\times P}$ be a known overcomplete basis parametrized by $\{\tilde\omega_1\ \cdots\ \tilde\omega_P\}$. More specifically, $\tilde\V$ can be expressed as
\begin{align}
  \tilde\V = \begin{bmatrix}
    \bfv(\tilde\omega_1) & \cdots & \bfv(\tilde\omega_P)
  \end{bmatrix}.
\end{align}
Note that $P$ should be much larger than the number of active frequencies $L$. Assuming a sufficiently dense grid, $\x$ can be approximated as
\begin{align}\label{s4:x2}
  \x \approx \tilde\V\tilde\x
\end{align}
where $\tilde\x \in\bC^P$ is $L$-sparse. Substituting \eqref{s4:x2} into \eqref{model1} yields
\begin{align}
  y_i \approx |\bfb_i^H\tilde\x|^2 + n_i,\ \forall i
\end{align}
where
\begin{align}
  \bfb_i = \tilde\V^H\a_i.
\end{align}
The problem of frequency estimation has been converted to sparse spectrum ($\tilde\x$) estimation. An ideal description of sparsity is the $\ell_0$-norm $||\x||_0$, i.e., the number of nonzero entries in $\x$. However, this yields a `doubly NP-hard' problem. In recent years, numerous approximations have been developed such as $\ell_1$ and $\ell_p\,(p<1)$ relaxations \cite{l1_1}-\cite{l1_2}, to replace the $\ell_0$-norm. For sparse B-FPP, we can use $\ell_1$ relaxation as follows
\begin{equation}
    \begin{aligned}
        \min_{\tilde\x,\s}\quad & ||\tilde\x||_1 + \lambda_1\sum_{i=1}^M s_i \\
        \text{s. t.}\quad
        &2\Re\{\z^H\B_i\tilde\x\} + s_i \geq \z^H\B_i\z + y_i-\epsilon \\
        & \tilde\x^H\B_i\tilde\x\leq y_i + \epsilon\\
        &s_i\geq 0,\ \forall i
        \end{aligned}\label{Q1_sparse}
\end{equation}
where $\B_i = \bfb_i\bfb_i^H\in\bC^{P\times P}$, $\tilde\x\in\bC^P$ and $\z\in\bC^P$. For sparse LS-FPP, we likewise have
\begin{equation}
  \begin{aligned}
    \min_{\tilde\x,\w,\s}\quad &||\w||^2_2 + \lambda_1||\tilde\x||_1 + \lambda_2\sum_{i=1}^M s_i\\
    \text{s. t.}\quad & 2\Re\{\z^H\B_i\tilde\x\}  + w_i + s_i\geq \z^H\B_i\z + y_i\\
    & \tilde\x^H\B_i\tilde\x + w_i \leq y_i, \\
    & s_i\geq 0, \forall i.
  \end{aligned}\label{Q2_sparse}
\end{equation}

\begin{remark}
	Similar to \textbf{Algorithm \ref{FPP}} and \textbf{Algorithm \ref{LS-FPP}} for `plain' phase retrieval, \eqref{Q1_sparse} and \eqref{Q2_sparse} can be solved repeatedly using the previously obtained $\tilde\x$ to obtain a new supporting point $\z$. Also note that sparse B-FPP and sparse LS-FPP are not limited to harmonic retrieval -- they are directly applicable to other cases where $\x$ admits a sparse representation in a known dictionary.
\end{remark}

\subsection{CRB for Harmonic Retrieval from Quadratic Measurements}
When $\x$ is modeled as a sum of a few harmonics, the CRB is associated to the unknown frequencies $\omega_\ell$ and complex amplitudes $\gamma_\ell$ rather than $\x$. The corresponding CRB is provided in the following theorem.
\begin{theorem}\label{theorem_crb_v}
  If $\x\in\bC^{N}$ is a superposition of $L$ Vandermonde vectors as in \eqref{s4:x}, the CRB is
  \begin{align}\label{crb_vdmd}
    \CRB_v = \F_v^\dagger
  \end{align}
  where
  \begin{align}
    \F_v = \frac{4}{\sigma_n^2}\G_v\G_v^T
  \end{align}
  with
  \begin{align}
    \G_v =& \begin{bmatrix}
      \Re\{\X^H\A_1\x\} & \cdots & \Re\{\X^H\A_M\x\} \\
      \Re\{\V^H\A_1\x\} & \cdots & \Re\{\V^H\A_M\x\} \\
      \Im\{\V^H\A_1\x\} & \cdots & \Im\{\V^H\A_M\x\}
    \end{bmatrix} \\
    \X =& \begin{bmatrix}
      \gamma_1\frac{\partial \bfv_1}{\partial\omega_1} & \cdots & \gamma_L\frac{\partial \bfv_L}{\partial\omega_L}
    \end{bmatrix} \\
    \V =& \begin{bmatrix}
          \bfv(\omega_1) & \cdots & \bfv(\omega_L)
        \end{bmatrix} \\
    \frac{\partial \bfv_\ell}{\partial\omega_\ell} =& \begin{bmatrix}
      je^{j\omega_\ell} & \cdots & jNe^{jN\omega_\ell}
    \end{bmatrix}^T.
  \end{align}
\end{theorem}
\begin{IEEEproof}
  See Appendix \ref{proof_of_vdmd}.
\end{IEEEproof}

Note that $\F_v$ is singular, and its rank deficit is equal to one. The proof is very similar to those in Appendices \ref{proof_of_rank1} and \ref{proof_of_rank1_2}, so it is omitted for brevity.

\section{Simulation Results}
We present simulations of the two proposed methods and compare them with WF \cite{C3}, GS \cite{AP:1}, PhaseLift \cite{C1}  and PhaseCut \cite{PC} in this section. The signal $\x=\exp{(j0.16\pi t)}, t=1,\cdots,N$, is deterministic and fixed throughout all Monte-Carlo trials. Furthermore, the SNR is defined as
\begin{align}
  \text{SNR} = \frac{\sum_{i=1}^M|\a_i^H\x|^4}{M\sigma_n^2}.
\end{align}
We consider two different types of measurements: 1) Gaussian measurements which are generated from a complex Gaussian distribution, i.e., the real and imaginary parts of each entry in $\a_i$ are generated from the normal distribution; 2)  masked Fourier measurements of the following form
\begin{equation}
\A^H =
\begin{bmatrix}
\F\D_1 \\ \vdots \\ \F\D_K
\end{bmatrix}
\end{equation}
where $K=M/N$, $\F$ is a $N\times N$ Fourier matrix and $\D_i$ is a $N\times N$ diagonal masking matrix with its diagonal entries independently generated by $b_1b_2$, where $b_1$ and $b_2$ are independent and distributed as \cite{C3}
\begin{equation}
b_1 = \left\{
\begin{aligned}
1\quad  &\text{with prob. 0.25} \\
-1\quad &\text{with prob. 0.25} \\
-j\quad &\text{with prob. 0.25} \\
j\quad  &\text{with prob. 0.25}
\end{aligned} \right.
\end{equation}
and
\begin{equation}
b_2 = \left\{
\begin{aligned}
\sqrt{2}/2\quad  &\text{with prob. 0.8} \\
\sqrt{3}\quad &\text{with prob. 0.2}.
\end{aligned} \right.
\end{equation}
The noise is assumed to be white Gaussian with mean zero and variance $\sigma_n^2$. The stopping criterion for B-FPP, LS-FPP, WF and GS is the relative improvement in the cost function value dropping below $10^{-7}$, i.e.,
\begin{align}
  \frac{||\y-|\A^H\x_k|^2||^2_2 - ||\y-|\A^H\x_{k-1}|^2||^2_2}{||\y-|\A^H\x_{k-1}|^2||^2_2} \leq 10^{-7}
\end{align}
or a limit on the maximum number of iterations being reached. This limit is set to 100, 100, 2000 and 2000, iterations for B-FPP, LS-FPP, WF and GS, respectively.

\subsection{CRB versus SNR}
As a first sanity check, Fig. 2 plots the CRB as a function of SNR for $M=2N,4N,\mathrm{and}\,8N$ for the complex-valued signal (top) and the real-valued signal (bottom), for $N=16$. It is seen that as predicted by Theorem \ref{crb_ap}, the bound on the standard deviation of the estimated $\x$ decreases as SNR increases. As expected, we also find that the CRB associated to a larger $M$ produces a smaller bound on the standard deviation, which validates our analytical results in \eqref{s2:crbm1}.

\begin{figure}[h]
\begin{center}
\subfigure[Amplitude]{\includegraphics[scale=0.4, trim=0 0 0 0]{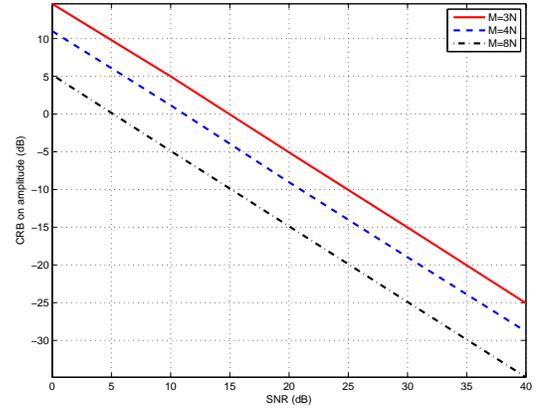}} 
\subfigure[Phase]{\includegraphics[scale=0.4,trim=0 0 0 0]{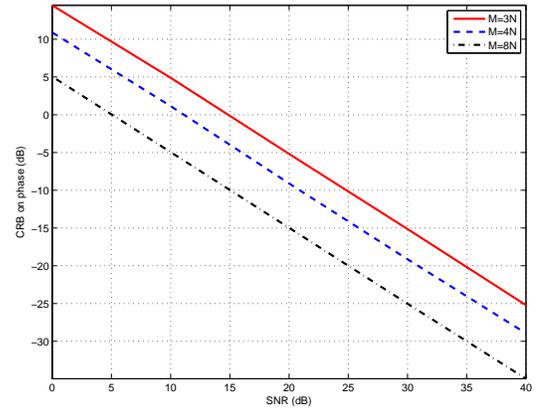}} 
\caption{CRB versus SNR for different $M$ with Gaussian measurements.}\vspace{-1.5em}
\end{center}
\end{figure}

\subsection{MSE Performance Comparison}
We now compare the performance of B-FPP and LS-FPP with PhaseLift, PhaseCut\footnote{PhaseCut works with $\sqrt{\y}$; $\y$ can have negative elements at low SNR, so we use $\Re\{\sqrt{\y}\}$ for PhaseCut. Also note that, due to the nonlinear transformation, noise will no longer be additive Gaussian for PhaseCut, which matches a different measurement model, namely $z_i = |{\bf a}_i^H \x|$.}, WF \cite{C3} and GS. For PhaseLift, PhaseCut, and WF, we use publicly available code\footnote{Downloaded from \url{http://www-bcf.usc.edu/~soltanol/PhaseRetrieval_CDP.zip}, \url{http://www.cmap.polytechnique.fr/scattering/code/phaserecovery.zip}, and \url{http://www-bcf.usc.edu/~soltanol/WFcode.html}, respectively.}. We use the LS version of PhaseLift that is appropriate for additive Gaussian noise. For B-FPP, $\epsilon$ is set equal to the standard deviation of the noise, for all our experiments.

To begin, let us illustrate the recovery performance of B-FPP and LS-FPP by means of example.  We set $N=16$, $M=64$ and SNR$=25$ dB. We consider two different initialization methods to start B-FPP, LS-FPP, WF and GS:
\begin{itemize}
  \item[1)] Spectrum initialization - picking the leading eigenvector of $\sum_{i=1}^My_i\a_i\a_i^H$ as an initial guess of $\x$;
  \item[2)] Gaussian random initialization - each element of the initial point is randomly generated from a complex Gaussian distribution with zero mean and unit variance.
\end{itemize}
Figs. 3 and 4 plot the histogram bar chart of 500 independent MSE samples where MSE is defined as
\begin{align}
  \text{MSE} =&\ 10\log_{10}\left(||\hat\x - \x||_2^2\right).
\end{align}
It is seen that for masked Fourier measurements, BS-FPP, LS-FPP, WF and GS perform very similarly and they outperform the PhaseLift and PhaseCut algorithms, since the latter frequently fail to find a rank-1 matrix. In the Gaussian measurement case, since both real and imaginary parts of each measurement vector are drawn from a standard normal distribution, $\mathbb{E}(\a_i\a_i^H)=2\I$. Therefore, the expected value of $\frac{1}{M}\sum_{i=1}^My_i\a_i\a_i^H$ is $2(\I+\x\x^H)$, and the top two eigenvectors of $(\I+\x\x^H)$ might be mixed together and the leading eigenvector will no longer be a good guess of $\x$ with a finite number of measurements. Due to this, we can see in Fig. 3(b) that the WF method (which is sensitive to the starting point), suffers from performance degradation. Furthermore, it is observed from Fig. 4 that by using random initialization, all the algorithms have more outages than the case in Fig. 3, and FPP-based methods are better than the others. Quantitative MSE results summarized in Table \ref{table_mse}, from which we can see that LS-FPP achieves the smallest variance in all the scenarios. Although it is seen from Figs. 3 and 4 that GS has as few outages as B- and LS-FPP, its MSE is still much larger than the latter methods. Note that the MSEs reported have been computed {\em after removing outages}, where we have defined MSE larger than 0 dB as an outage. The CRB is an averaged result over 500 Monte-Carlo tests and is computed via Theorem \ref{theorem_crb_c}. Furthermore, it is seen in Table II that for masked Fourier measurements the outage percentage of LS-FPP is slightly larger than that of WF; while for Gaussian measurements, FPP-based methods are much better than WF and GS. It is interesting that, although the MSEs of PhaseLift and PhaseCut are not as good, the two relaxation-based methods still do very well in terms of avoiding outages.

\noindent {\it Remark:}
These results suggest using the principal eigenvector of SDR to initialize FPP, and indeed this further reduces the number of outages, as well as the number of outer iterations in B-FPP and LS-FPP. The drawback is that as the size of $N=\text{length}(\x)$ becomes larger, SDR quickly becomes the complexity bottleneck, since it lifts the problem to a much higher-dimensional $O(N^2)$ space. Still, using SDR for initialization is well worth the effort for smaller $N$, as the overall complexity is still of the same order as that of FPP {\em per se}. This is never the case for WF and GS, which are relatively lightweight algorithms whose computational cost is always dominated by SDR.

\begin{figure}[ht]
\begin{center}
\subfigure[Masked Fourier measurements]{\includegraphics[scale=0.4,trim=0 0 0 0]{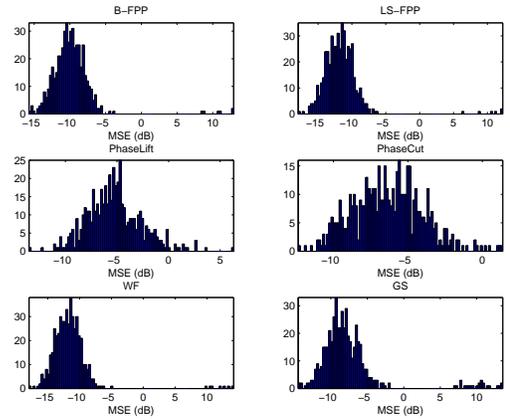}}
\subfigure[Gaussian measurements]{\includegraphics[scale=0.4, trim=0 0 0 0]{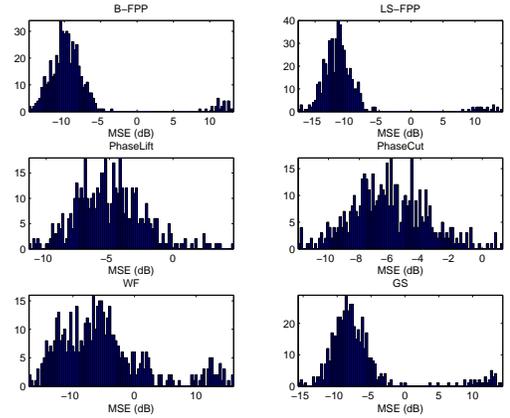}}
\caption{Signal recovery performance comparison with spectrum initialization.}\vspace{-1em}
\end{center}
\end{figure}

\begin{figure}[ht]
\begin{center}
\subfigure[Masked Fourier measurements]{\includegraphics[scale=0.4,trim=0 0 0 0]{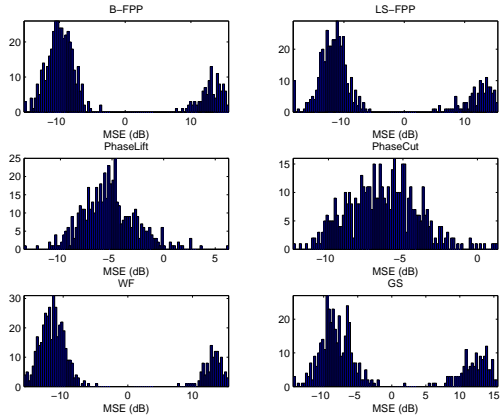}}
\subfigure[Gaussian measurements]{\includegraphics[scale=0.4, trim=0 0 0 0]{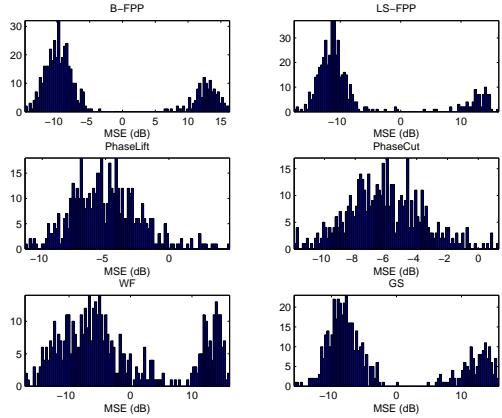}}
\caption{Signal recovery performance comparison with random initialization.}\vspace{-1.5em}
\end{center}
\end{figure}

\begin{table*}
\begin{center}
    \caption{Averaged MSE and CRB (both in dB) after removing outages}\label{table_mse}\vspace{-0.5em}
  \renewcommand{\arraystretch}{1.5}
  \begin{tabular}{ l | c | c | c | c | c | c | c | r }
    \hline
    \multicolumn{2}{c|}{Setting} & CRB & B-FPP & LS-FPP & PhaseLift & PhaseCut & WF & GS \\ [0.7ex]
    \hline
     \multirow{ 2}{*}{Masked Fourier} & Spec. Init. &  -11.4268 & -9.6536 & \textbf{-11.4208} & -4.8509 & -5.8273 & -11.4137 & -7.9174 \\  \cline{2-9}
     & Rand. Init. & -11.4268 & -9.5166 & \textbf{-11.2285} & -4.8509 & -5.8273 &  -11.2711 & -7.6803 \\
     \hline
     \multirow{ 2}{*}{Gaussian Meas.} & Spec. Init. &  -11.0616 & -9.3672 & \textbf{-11.0596} & -4.6289 & -5.5361 & -6.1576 & -7.5980 \\  \cline{2-9}
     & Rand. Init. & -11.0616 & -9.3588 &  -\textbf{10.5681} & -4.6289 & -5.5361 & -6.0553 & -7.5412 \\
     \hline
  \end{tabular}
\end{center}
\end{table*}

\begin{table*}
\begin{center}
    \caption{Outage percentages}\label{table_no_outages}\vspace{-0.5em}
  \renewcommand{\arraystretch}{1.5}
  \begin{tabular}{ l | c | c | c | c | c | c | r }
    \hline
    \multicolumn{2}{c|}{Setting} & B-FPP & LS-FPP & PhaseLift & PhaseCut & WF & GS \\ [0.7ex]
    \hline
     \multirow{ 2}{*}{Masked Fourier} & Spec. Init. & 2.6\% & 2.2\% & 2.8\% & 1\% & 1.8\% & 4.6\% \\  \cline{2-8}
     & Rand. Init. & 22.2\% & 24\% & 2.8\% & 1\%  &  23.2\% & 28.6\% \\
     \hline
     \multirow{ 2}{*}{Gaussian Meas.} & Spec. Init. & 4.8\% & 4.4\% & 3.8\% & 1.2\% & 17.8\% & 7.4\% \\  \cline{2-8}
     & Rand. Init. & 26\% & 16.8\% & 3.8\% & 1.2\% & 31.4\% & 27.8\% \\
     \hline
  \end{tabular}
\end{center}
\end{table*}

Next, we compare the MSE performance as a function of SNR, using $N=16$, $M=128$, and 200 Monte-Carlo trials. 
\begin{align*}
  &\text{MSE on amplitude} = 10\log_{10}\left(\frac{1}{200}\sum_{i=1}^M\left|\left|\, |\hat\x|_i - |\x|\, \right|\right|_2^2\right) \\
  &\text{MSE on phase} = 10\log_{10}\left(\frac{1}{200}\sum_{i=1}^M\left|\left|\, \angle(\hat\x)_i - \angle(\x)\, \right|\right|_2^2\right)
\end{align*}
where $\angle(\cdot)$ takes the phase of its argument.
The CRB in Theorem \ref{crb_ap} is also included as a benchmark. Fig. 5 depicts the MSE results for masked Fourier measurements, from which we observe that LS-FPP and WF followed by B-FPP achieve the best performance and all of them outperform PhaseLift, PhaseCut and GS when SNR is higher than 10 dB.
In Fig. 6, GS and WF exhibit relative high MSE in the high SNR regime, which is mainly caused by occasional outages (we noted that GS and WF produce three or four outages during the 200 Monte-Carlo trials, at SNR $>30$ dB). When SNR $\leq5$ dB, there is no MSE value reported for WF because WF frequently returns NaN (\emph{not a number}). The reason is that the noise variance is commensurate to the useful signal power and the eigenvalues of $\sum_{i=1}^M\a_i\a_i^H$ are of the same order, thus the leading eigenvector is no longer useful as initialization. Note that WF, LS PhaseLift, and LS-FPP actually attempt to solve the same problem formulation here, however only LS-FPP is insensitive to initialization and competitive in terms of statistical efficiency in this scenario.

\begin{figure}[h]
	\begin{center}
		\subfigure[MSE on amplitude]{\includegraphics[scale=0.4, trim=0 0 0 0]{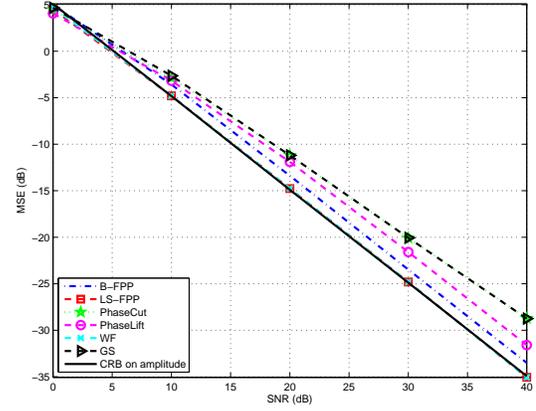}}
		\subfigure[MSE on phase]{\includegraphics[scale=0.4,trim=0 0 0 0]{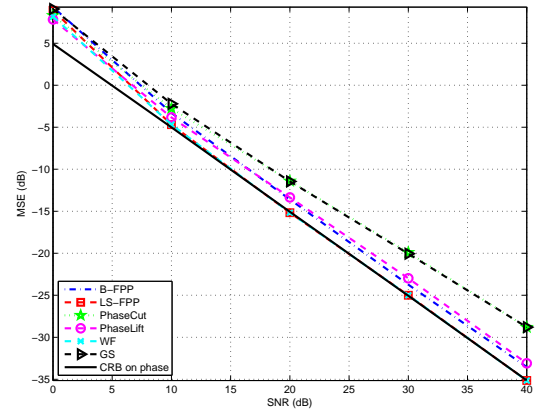}}
		\caption{Performance comparison with masked Fourier measurements.}\vspace{-1.5em}
	\end{center}
\end{figure}

\begin{figure}[h]
	\begin{center}
		\subfigure[MSE on amplitude]{\includegraphics[scale=0.4, trim=0 0 0 0]{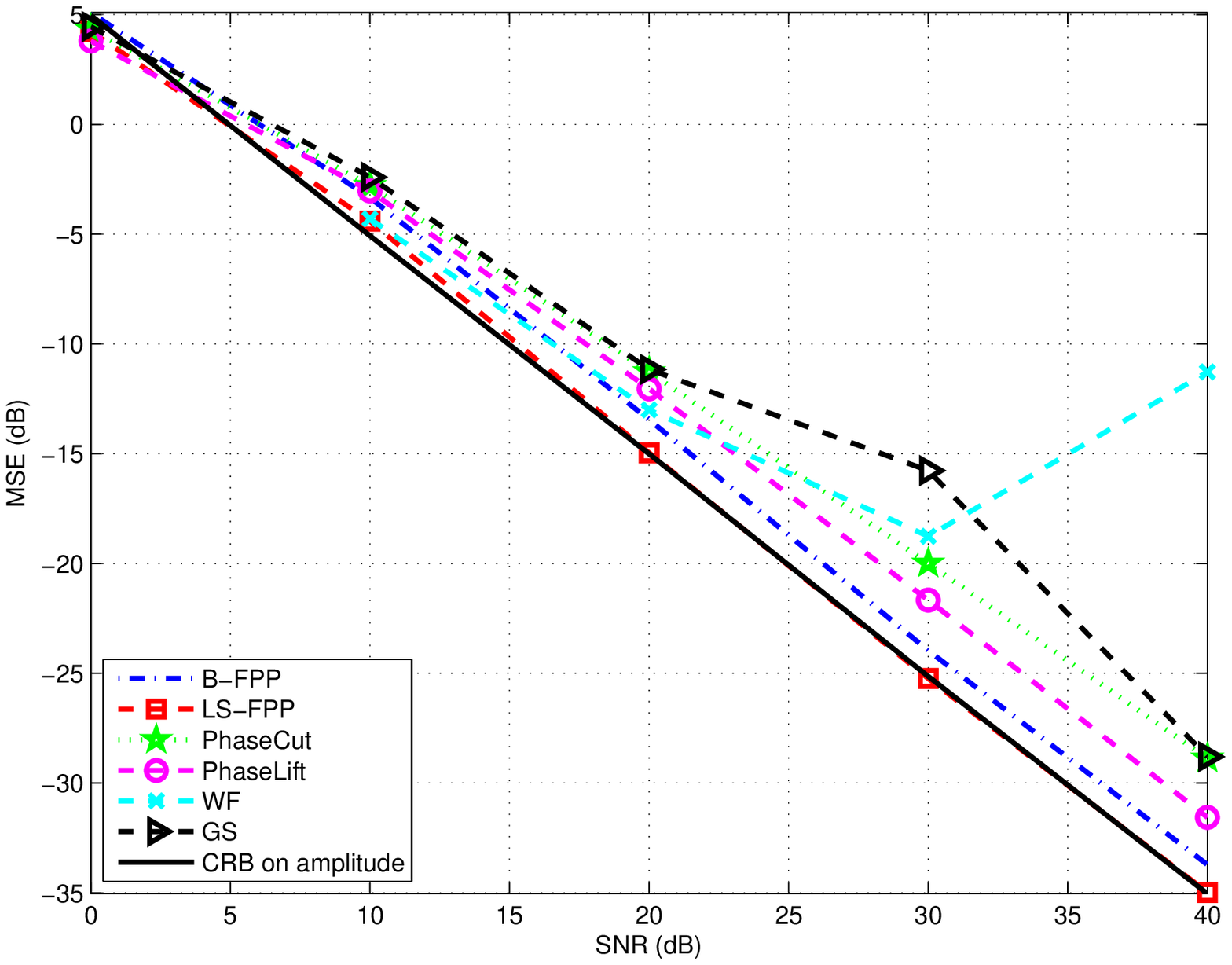}}
		\subfigure[MSE on phase]{\includegraphics[scale=0.4,trim=0 0 0 0]{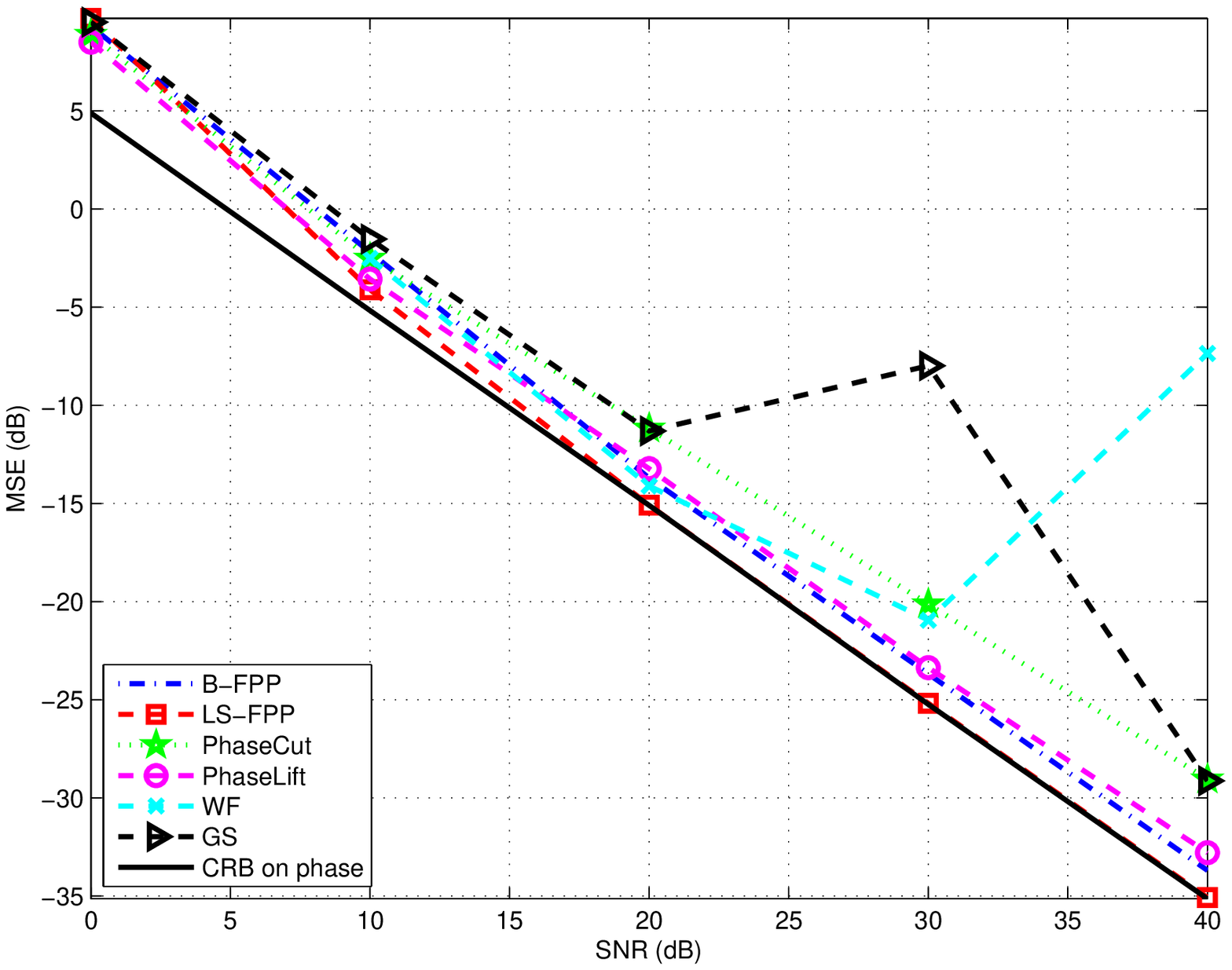}}
		\caption{Performance comparison with Gaussian measurements.}\vspace{-1em}
	\end{center}
\end{figure}

\subsection{Performance Comparison for Harmonic Retrieval from Rank-one Quadratic Measurements}
\begin{figure}[h]
	\centering
	\includegraphics[width=7.2cm]{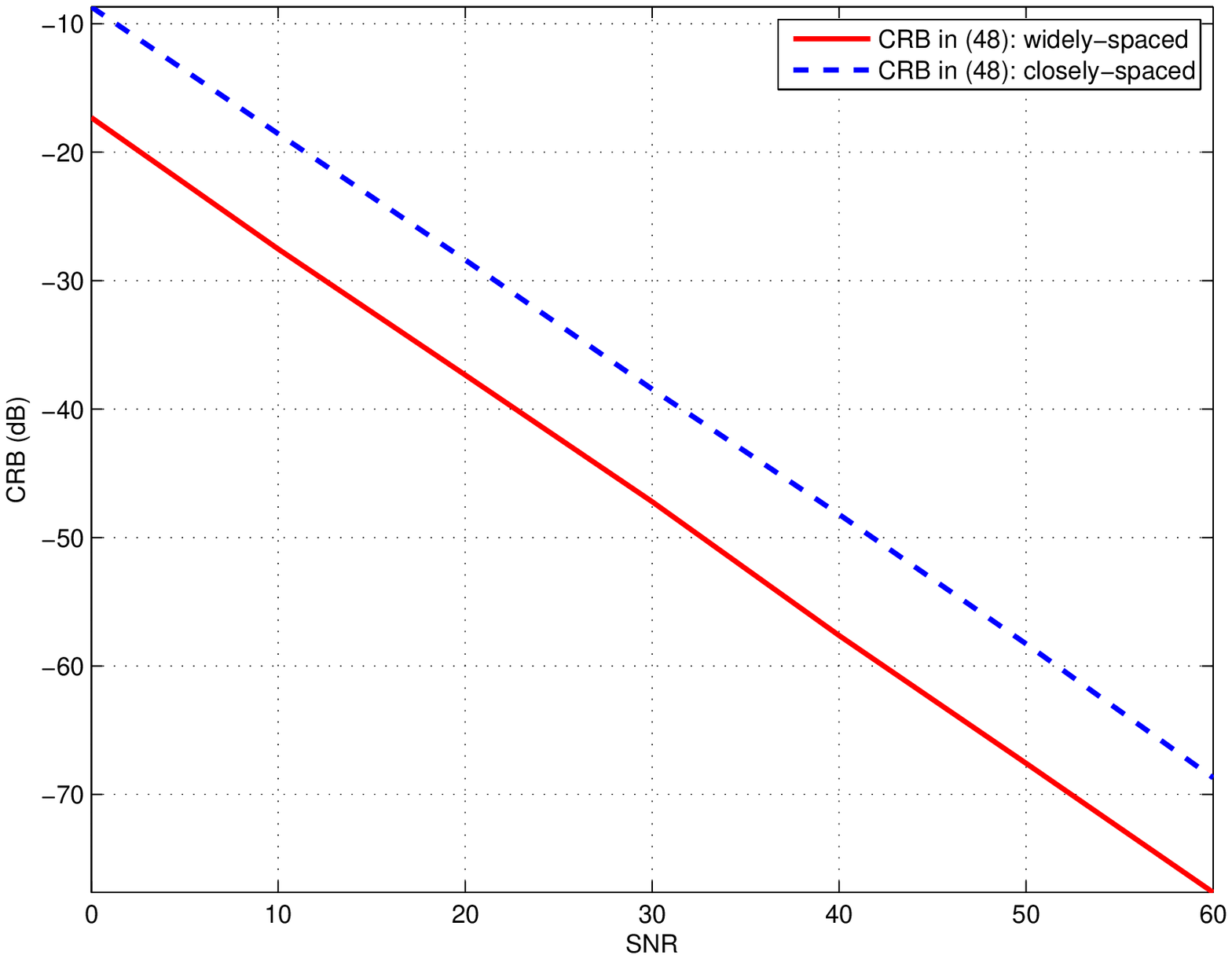}\\
	\caption{CRB versus SNR for harmonic retrieval from quadratic measurements.}\label{CRB_Vandermonde_x}\vspace{-1em}
\end{figure}

\begin{figure}[h]
	\centering
	\includegraphics[width=7.2cm]{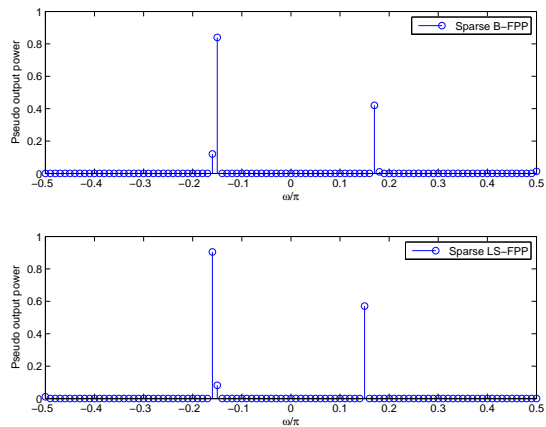}\\
	\caption{Signal recovery for harmonic retrieval from quadratic measurements.}\label{Recovery_V_x}\vspace{-1em}
\end{figure}

We consider a scenario where $\x$ has the form of a 1-D harmonic model. Assume that there are two frequencies contained in $\x$, i.e.,
$$\x = \bfv(\omega_1)+\bfv(\omega_2).$$
We study the CRB in \eqref{crb_vdmd} as a function of SNR. In this example, we assume that $N=8$ and $M=40$. Fig. \ref{CRB_Vandermonde_x} plots two CRB curves corresponding to widely-spaced frequencies ($\omega_1=-0.15\pi$ and $\omega_2=0.15\pi$) and closely-spaced frequencies ($\omega_1=-0.05\pi$ and $\omega_2=0.05\pi$). As expected, the CRB for closely-spaced frequencies is larger than that for widely-spaced ones.
Fig. \ref{Recovery_V_x} plots the pseudo power spectra, i.e., $\tilde \x$, obtained by sparse B-FPP and sparse LS-FPP. In this example, the parameters are $\omega_1=-0.16\pi$, $\omega_2=0.16\pi$, $N=8$, $M=16$ and SNR $=30$ dB. The dictionary is of length 51, obtained by uniformly sampling the $[-\pi/2, \pi/2]$ frequency sector. It is observed from Fig. \ref{Recovery_V_x} that sparse LS-FPP has two distinct peaks around the true $\omega$, while sparse B-FPP has a small bias on the estimate of $\omega_2$.

\section{Conclusions}
The problem of phase retrieval has been revisited from a non-convex QCQP point of view. Building upon recent work on feasible point pursuit for non-convex QCQP problems, two novel algorithms were developed for phase retrieval from noisy measurements: B-FPP and LS-FPP. B-FPP is designed for uniform additive noise, such as quantization noise introduced by high-resolution uniform quantization. LS-FPP is matched to white Gaussian noise that is added after taking the magnitude squared of the linear measurements, such as analog transmission noise. For the latter model, the Cram\'er-Rao bound was also derived and studied. Simulations suggest that B-FPP and LS-FPP attain state-of-art performance, and LS-FPP outperforms all earlier methods and comes very close to the CRB under certain conditions (depending on the SNR, and the type and number of measurements relative to the signal dimension). It was also shown that what apparently hurts the average performance of some of the most competitive algorithms is outages, even when they are rare. LS-FPP exhibits the best outage performance among all algorithms considered, including WF, which seems to be quite sensitive to outages, especially for systematic (as opposed to i.i.d. Gaussian) measurement vectors, which throw off its initialization. Variations of B-FPP and LS-FPP (and the corresponding CRB) for harmonic retrieval from rank-1 quadratic measurements were also developed and illustrated in simulations. The drawback of B-FPP and LS-FPP is their relatively high computational complexity, especially compared to WF. Ways of bringing down this complexity are currently under investigation.

\appendices

\section{Proof of Theorem \ref{theorem_crb_c}}\label{App_proof_cmplx}
The CRB states that the variance of any unbiased estimator is at least as high as the inverse of the FIM. To determine the CRB, we should first calculate the FIM and then take its inverse. The likelihood function for the data model for complex $\x$ is
\begin{align}\label{likelihood}
    p(\y;\x) = \prod_{i=1}^M\frac{1}{\sqrt{2\pi\sigma_n^2}}\exp\left\{-\frac{(y_i-\x^H\A_i\x)^2}{2\sigma_n^2}\right\}.
\end{align}
Hence, the log-likelihood function can be written as
\begin{align}
    \ln p(\y;\x) = -\frac{M}{2}\ln(2\pi\sigma_n^2) - \frac{1}{2\sigma_n^2}\sum_{i=1}^M(y_i-\x^H\A_i\x)^2.
\end{align}
The vector of unknown parameters for complex $\x$ is
\begin{align}\label{A2:beta}
  \bbe = [\ b_1\ \cdots\ b_N, \theta_1\ \cdots\ \theta_N]^T
\end{align}
where $b_i$ and $\theta_i$ are the amplitude and phase of $x_i$, i.e.,
\begin{equation}
	x_i=b_ie^{j\theta_i}.
\end{equation}
Thus, the FIM can be expressed as
\begin{align}\label{CRB:F}
      \F =
    \begin{bmatrix}
     \F_{bb} & \F_{b\theta} \\
     \F_{\theta b}  & \F_{\theta\theta}
    \end{bmatrix}
\end{align}
where the $(m,n)$ entry of the FIM is given by
\begin{align}\label{F}
    [  \F]_{m,n} = -\bE\left[\frac{\partial^2 \ln p(\y;\x)}{\partial \bbe_m\partial \bbe_n}\right]
\end{align}
and
\begin{align}
  [\F_{bb}]_{m,n} =&\ -\bE\left[\frac{\partial^2 \ln p(\y;\x)}{\partial b_m\partial b_n}\right] \\
  [\F_{\theta\theta}]_{m,n} =&\ -\bE\left[\frac{\partial^2 \ln p(\y;\x)}{\partial \theta_m\partial \theta_n}\right]\\
  [\F_{\theta b}]_{m,n} =&\ -\bE\left[\frac{\partial^2 \ln p(\y;\x)}{\partial \theta_m\partial b_n}\right] \\
  [\F_{b\theta}]_{m,n} =&\ -\bE\left[\frac{\partial^2 \ln p(\y;\x)}{\partial b_m\partial \theta_n}\right] .
\end{align}

The second-order derivative of $\ln p(\y;\x)$ is
\begin{align}\label{CRB:1}
	\frac{\partial^2\ln p(\y;\x)}{\partial \bbe_m\partial \bbe_n} = \frac{1}{\sigma_n^2}\sum_{i=1}^M&\
	\left( (y_i-\x^H\A_i\x)\frac{\partial^2\x^H\A_i\x}{\partial \bbe_m\partial \bbe_n}\right. \notag\\
	&\qquad\left.- \frac{\partial\x^H\A_i\x}{\partial \bbe_m}\frac{\partial\x^H\A_i\x}{\partial \bbe_n}\right).
\end{align}
Taking the expectation of both sides of \eqref{CRB:1} produces that
\begin{align}\label{CRB:2}
	\bE\left[\frac{\partial^2\ln p(\y;\x)}{\partial \bbe_m\partial \bbe_n}\right] = -\frac{1}{\sigma_n^2}\sum_{i=1}^{M}\frac{\partial\x^H\A_i\x}{\partial \bbe_m}\frac{\partial\x^H\A_i\x}{\partial \bbe_n}
\end{align}
where $\bE[y_i-\x^H\A_i\x] = 0$. Now,
\begin{align}\label{CRB:3}
	\frac{\partial \x^H\A_i\x}{\partial \theta_m} =&\ -jx_m^*\A(m,:)\x + jx_m\x^H\A(:,m) \notag \\
	=&\ 2\Re\{-jx_m^*\A(m,:)\x \}
\end{align}
\begin{align}\label{CRB:4}
	\frac{\partial \x^H\A_i\x}{\partial b_m} =&\ e^{-j\theta_m}\A(m,:)\x + e^{j\theta_m}\x^H\A(:,m) \notag \\
	=&\ 2\Re\{e^{-j\theta_m}\A(m,:)\x \}
\end{align}
where $(\cdot)^*$ is the conjugate, and $\A(i,:)$ and $\A(:,i)$ stand for the $i$th row and column of $\A $, respectively.
Thus, by substituting \eqref{CRB:3} and \eqref{CRB:4} into \eqref{CRB:2},
after some matrix manipulations, we obtain the matrix form of the sub-FIMs as
\begin{align}
    \F_{bb} =&\ \frac{4}{\sigma_n^2}\Re\{\diag(e^{-j\bth})\A^H\diag(\A\x)\}\notag\\ &\ \times\Re\{\diag(e^{-j\bth})\A^H\diag(\A\x)\}^T \label{F111} \\
	\F_{\theta\theta} =&\ -\frac{4}{\sigma_n^2}\Re\{\diag(\x^*)\A^H\diag(\A\x)\}\notag\\ &\ \times\Re\{\diag(\x^*)\A^H\diag(\A\x)\}^T \label{F222}\\
	\F_{\theta b} =&\ -\frac{4}{\sigma_n^2}\Re\{j\diag(\x^*)\A^H\diag(\A\x)\}\notag\\ &\ \times\Re\{\diag(e^{-j\bth})\A^H\diag(\A\x)\}^T \label{333}\\
	\F_{b\theta} =&\ \frac{4}{\sigma_n^2}\F_{\theta b}^T \label{F444}
\end{align}
where $\bth=[\theta_1\ \cdots\ \theta_N]^T$.
Inserting \eqref{F111} to \eqref{F444} into \eqref{F} produces the whole FIM
\begin{align}
	  \F = \frac{4}{\sigma_n^2}  \G  \G^T
\end{align}
where
\begin{align}
	   \G =
	\begin{bmatrix}
	\Re\left\{\diag(e^{-j\bth})\A\diag(\A^H\x)\right\}\\
	\Im\left\{\diag(\x^*)\A\diag(\A^H\x)\right\}
	\end{bmatrix}.
\end{align}
Using block matrix inverse formula, the CRB associated to the phase and amplitude can be expressed as
\begin{align}
	\text{CRB}_\theta^\dagger =&\ \F_{\theta\theta} - \F_{\theta b}\F_{bb}^{-1}\F_{b\theta} \\
	\text{CRB}_b^{-1} =&\ \F_{bb} - \F_{b\theta} \F_{\theta\theta}^\dagger\F_{\theta b}.
\end{align}

\section{Rank-1 deficiency of $\F$}\label{proof_of_rank1}
To show that $\F$ is rank-1 deficient, it suffices to find a non-zero vector ${\bf v}$ such that $\F{\bf v}={\bf 0}$.

Denote ${\bf v}\in\R^{2N}$ as $[~{\bf v}_1^T~{\bf v}_2^T~]^T$, then
\begin{align}
	\G^T{\bf v} =&\ \Re\{\diag(e^{-j\bth})\A\text{diag}(\A^H\x)\}^T{\bf v}_1 \notag\\
	&\ + \Im\{\diag(\x^*)\A\text{diag}(\A^H\x)\}^T{\bf v}_2 \notag\\
	=&\ \Re\{\diag(\x^*)\A\text{diag}(\A^H\x)\}^T\tilde{\bfv}_1 \notag\\
	&\ + \Im\{\diag(\x^*)\A\text{diag}(\A^H\x)\}^T{\bf v}_2
\end{align}
where $\tilde\bfv_1=\diag(|\x|)^{-1}{\bf v}_1$.
Now let ${\bf u}= {\bf v}_2 + j\tilde{\bf v}_1$, then
\begin{align}
	\G^T{\bf v} & = \Im\left\{ \left(\diag(\x^*)\A\text{diag}(\A^H\x)\right)^H{\bf u} \right\} \\
	& = \Im\left\{ (\A^H\x)^*\odot\left(\A^H\diag(\x){\bf u}\right) \right\}.
\end{align}
Let ${\bf u}=\alpha\boldsymbol{1}_N\in\bR^N$, for any $\alpha \neq 0$; then
\begin{align}
	\G^T{\bf v} = \Im\left\{ \alpha\left|\A^H\x\right|^2 \right\} = \0.
\end{align}
This means that the direction ${\bf v} = \alpha[~\zero_N^T~\boldsymbol{1}_N^T~]^T$, which is non-zero, lies in the null space of $\G$, thus also in the null space of $\F$.
Moreover, suppose the vector $\A^H\x$ does not contain any zero elements, which is true almost surely. To find another null space of $\G$ would require the vector $\A^H\diag(\x){\bf u}$ to be the all zero vector. Such a vector ${\bf u}$ does not exist almost surely, for example if $\A$ is a random Gaussian matrix. This means $\F$ is rank-1 deficient almost surely.

\section{Rank-1 deficiency of $\F_c$}\label{proof_of_rank1_2}
Denote ${\bf v}\in\R^{2N}$ as $[~{\bf v}_1^T~{\bf v}_2^T~]^T$, then
$$
\G_c^T{\bf v} = \Re\{\A\text{diag}(\A^H\x)\}^T{\bf v}_1 + \Im\{\A\text{diag}(\A^H\x)\}^T{\bf v}_2.
$$
Now let ${\bf u}={\bf v}_1 + j{\bf v}_2$, then
\begin{align*}
\G_c^T{\bf v} & = \Re\left\{ \left(\A\text{diag}(\A^H\x)\right)^H{\bf u} \right\} \\
& = \Re\left\{ (\A^H\x)^*\odot(\A^H{\bf u}) \right\}
\end{align*}
Let ${\bf u}=j\x$, we have
$$
\G_c^T{\bf v} = \Re\left\{ j\left|\A^H\x\right|^2 \right\} = \0.
$$
This means the direction ${\bf v} = [~-\Im\{\x\}^T~\Re\{\x\}^T~]^T$, which is non-zero, lies in the null space of $\G_c$, thus also in the null space of $\F_c$.

Moreover, suppose the vector $\A^H\x$ does not contain any zero elements, which is true almost surely. To find another null space of $\G_c$ would require the vector $\A^H{\bf u}$ to be the all zero vector. Such a vector ${\bf u}$ does not exist almost surely, for example if $\A$ is a random Gaussian matrix. This means $\F_c$ is rank-1 deficient almost surely. It is also interesting to observe that for the Fisher information matrix with respect to an arbitrary complex signal $\x$, the direction $j\x$ is always in its null space.

\section{Proof of Full Rank of $\F_r$}\label{proof_of_fullrank_Fr}
Similarly, to show that $\F_r$ is full rank, it suffices to show that there does not exist a non-zero vector ${\bf v}\in\R^N$ such that $\F_r{\bf v} = \0$, or equivalently $\G_r^T{\bf v}=\0$. Again we have that
$$
\G_r^T{\bf v} = \Re\left\{ (\A^H\x)^*\odot(\A^H{\bf v}) \right\}.
$$
However, we are not allowed to choose ${\bf v}=j\x$ to make this product zero, because ${\bf v}$ can only be real. Assume $\A^H\x$ does not contain any zero elements, which is true almost surely, we must find a ${\bf v}$ such that $\A^H{\bf v}=\0$, which cannot happen almost surely. Therefore, $\F_r$ is full rank almost surely.

\section{Proof of Theorem \ref{theorem_crb_v}}\label{proof_of_vdmd}
The likelihood function for $\x$ equal to a sum-of-harmonics as in \eqref{s4:x} has the same expression as \eqref{likelihood}. However, the parameter vector contains the $L$ unknown frequencies and the real and imaginary parts of the $L$ unknown complex amplitudes $\{\gamma_1\ \cdots\ \gamma_L\}$:
\begin{align}
  \bal = [\omega_1\ \cdots\ \omega_L, \Re\{\gamma_1\}\ \cdots \ \Re\{\gamma_L\}, \Im\{\gamma_1\}\ \cdots\ \Im\{\gamma_L\}]^T.
\end{align}
The FIM associated to $\bal$ is expressed as
\begin{align}
  \F_v =
  \begin{bmatrix}
    \F_{\omega\omega} & \F_{\omega\Re\{\gamma\}} & \F_{\omega\Im\{\gamma\}} \\
    \F_{\Re\{\gamma\}\omega} & \F_{\Re\{\gamma\}\Re\{\gamma\}} & \F_{\Re\{\gamma\}\Im\{\gamma\}} \\
    \F_{\Im\{\gamma\}\omega} & \F_{\Im\{\gamma\}\Re\{\gamma\}} & \F_{\Im\{\gamma\}\Im\{\gamma\}}
  \end{bmatrix}\label{fv}
\end{align}
where
\begin{align}
  \F_{\Re\{\gamma\}\omega} =&\ \F_{\omega\Re\{\gamma\}}^T \\
  \F_{\Im\{\gamma\}\omega} =&\ \F_{\omega\Im\{\gamma\}}^T \\
  \F_{\Re\{\gamma\}\Im\{\gamma\}} =&\ \F_{\Im\{\gamma\}\Re\{\gamma\}}^T.
\end{align}
Therefore, we only need to calculate the upper triangular part of $\F_v$.
\begin{align}\label{CRB:vx}
    \frac{\partial\ln p(\y;\x)}{\partial\bal_m} = \frac{1}{\sigma_n^2}\sum_{i=1}^M\left( (y_i-\x^H\A_i\x)\frac{\partial\x^H\A_i\x}{\partial\bal_m} \right).
\end{align}
Let us first compute
\begin{align}
  \frac{\partial\x^H\A_i\x}{\partial\omega_m} =&\ \gamma_m^*\left(\frac{\partial\bfv_m}{\partial\omega_m}\right)^H\A_i\x + \gamma_m\x^H\A_i\frac{\partial\bfv_m}{\partial\omega_m} \notag\\
     =&\ 2\Re\left\{ \gamma_m^*\left(\frac{\partial\bfv_m}{\partial\omega_m}\right)^H\A_i\x \right\}
\end{align}
where
\begin{align}
  \frac{\partial\bfv_m}{\partial\omega_m} = \begin{bmatrix}
    je^{j\omega_m} & \cdots & jNe^{jN\omega_m}
  \end{bmatrix}^T.
\end{align}
In the sequel, we compute
\begin{align}\label{CRB:vx2}
    \frac{\partial^2\ln p(\y;\x)}{\partial\bom_m\partial\bom_n} =&\ \frac{1}{\sigma_n^2}\sum_{i=1}^M\left( (y_i-\x^H\A_i\x)\frac{\partial^2\x^H\A_i\x}{\partial\omega_m\partial\omega_n} \right. \notag\\
    &\qquad\quad\qquad\left. - \frac{\partial\x^H\A_i\x}{\partial\omega_m}\frac{\partial\x^H\A_i\x}{\partial\omega_n} \right).
\end{align}
To obtain \eqref{CRB:vx2}, we consider two cases to calculate the value of $\frac{\partial^2\x^H\A_i\x}{\partial\omega_m\partial\omega_n}$. If $m\neq n$,
\begin{align}
  \frac{\partial^2\x^H\A_i\x}{\partial\omega_m\partial\omega_n} = 2\Re\left\{\gamma_m^*\gamma_n \left(\frac{\partial\bfv_m}{\partial\omega_m}\right)^H\A_i\frac{\partial\bfv_n}{\partial\omega_n}\right\}.
\end{align}
If $m=n$,
\begin{align}
  \frac{\partial^2\x^H\A_i\x}{\partial\omega_m\partial\omega_n} =&\ 2|\gamma_m|^2\left(\frac{\partial\bfv_m}{\partial\omega_m}\right)^H\A_i\frac{\partial\bfv_m}{\partial\omega_m} \notag\\
  &+ 2\Re\left\{\gamma_m^*\left(\frac{\partial^2\bfv_m}{\partial\omega_m^2}\right)^H\A_i\x\right\}
\end{align}
where
\begin{align}
  \frac{\partial^2\bfv_m}{\partial\omega_m^2} = -\begin{bmatrix}
    e^{j\omega_m} & \cdots & N^2e^{jN\omega_m}
  \end{bmatrix}^T.
\end{align}
Taking the expectation of both sides of \eqref{CRB:vx2} yields
\begin{align}
  [\F_{\omega\omega}]_{m,n} =&\ \frac{4}{\sigma_n^2} \sum_{i=1}^M\Re\left\{ \gamma_m^*\left(\frac{\partial\bfv_m}{\partial\omega_m}\right)^H\A_i\x \right\}\notag\\
        &\ \times\Re\left\{ \gamma_n^*\left(\frac{\partial\bfv_n}{\partial\omega_n}\right)^H\A_i\x \right\}. \label{Fv11}
\end{align}
We next compute $\F_{\omega\Re\{\gamma\}}$. Here, we point out that $\alpha_m$ corresponds to frequencies while $\alpha_n$ corresponds to the real parts of the amplitudes.
\begin{align}
  \frac{\partial\x^H\A_i\x}{\partial\Re\{\gamma_n\}} = \bfv_n^H\A_i\x + \x^H\A_i\bfv_n = 2\Re\{\bfv_n^H\A_i\x\}.
\end{align}
Since the expected value of $(y_i-\x^H\A_i\x)$ is zero, we directly obtain
\begin{align}
  [\F_{\omega\Re\{\gamma\}}]_{m,n} =&\ \frac{4}{\sigma_n^2} \sum_{i=1}^M\Re\left\{ \gamma_m^*\left(\frac{\partial\bfv_m}{\partial\omega_m}\right)^H\A_i\x \right\}\notag\\
        &\ \times\Re\left\{ \bfv_n^H\A_i\x \right\} \label{Fv12}.
\end{align}
In a similar manner, 
\begin{align}
  \frac{\partial\x^H\A_i\x}{\partial\Im\{\gamma_n\}} = 2\Im\{\bfv_n^H\A_i\x\}
\end{align}
which results in the following formula for $\F_{\omega\Im\{\gamma\}}$
\begin{align}
  [\F_{\omega\Im\{\gamma\}}]_{m,n} =&\ \frac{4}{\sigma_n^2} \sum_{i=1}^M\Re\left\{ \gamma_m^*\left(\frac{\partial\bfv_m}{\partial\omega_m}\right)^H\A_i\x \right\}\notag\\
        &\ \times\Im\left\{ \bfv_n^H\A_i\x \right\} \label{Fv13}.
\end{align}
At this point, the expressions for the (m,n)th element of $\F_{\Re\{\gamma\}\Re\{\gamma\}}$ and $\F_{\Im\{\gamma\}\Im\{\gamma\}}$ can be easily derived
\begin{align}
  \left[\F_{\Re\{\gamma\}\Re\{\gamma\}}\right]_{m,n} =&\ \frac{4}{\sigma_n^2} \sum_{i=1}^M \Re\left\{ \bfv_m^H\A_i\x \right\} \Re\left\{ \bfv_n^H\A_i\x \right\} \label{Fv22} \\
  \left[\F_{\Im\{\gamma\}\Im\{\gamma\}}\right]_{m,n} =&\ \frac{4}{\sigma_n^2} \sum_{i=1}^M \Im\left\{ \bfv_m^H\A_i\x \right\} \Im\left\{ \bfv_n^H\A_i\x \right\} \label{Fv33}.
\end{align}
Substituting \eqref{Fv11}, \eqref{Fv12}, \eqref{Fv13}, \eqref{Fv22} and \eqref{Fv33} into \eqref{fv}, after some matrix manipulations, we have
\begin{align}
  \F_v = \frac{4}{\sigma_n^2} \G_v\G_v^T
\end{align}
where
\begin{align}
  \G_v =&\ \begin{bmatrix}
      \Re\{\X^H\A_1\x\} & \cdots & \Re\{\X^H\A_M\x\} \\
      \Re\{\V^H\A_1\x\} & \cdots & \Re\{\V^H\A_M\x\} \\
      \Im\{\V^H\A_1\x\} & \cdots & \Im\{\V^H\A_M\x\}
    \end{bmatrix} \\
  \X =&\ \begin{bmatrix}
    \gamma_1\frac{\partial\bfv_1}{\partial\omega_1} & \cdots & \gamma_L\frac{\partial\bfv_L}{\partial\omega_L}
  \end{bmatrix} \\
  \V =&\ [\bfv(\omega_1)\ \cdots\ \bfv(\omega_L)].
\end{align}
Note that using a similar proof as for the rank-1 deficiency property of the FIM in \eqref{FIM_c}, it can be easily shown that $\F_v$ is also rank-1 deficient. As a result, the CRB for sum-of-harmonics $\x$ is computed using the pseudo-inverse of $\F_v$.

\end{document}